\documentclass[12pt]{article}
\usepackage{psfig}
\usepackage{graphicx}

\begin{document}

\begin{center}

{\Huge\bf Bulk Motions of Spiral Galaxies in the $z = 0.03$ Volume}

\bigskip

{\large Yu.N.Kudrya$^1$,
V.E.Karachentseva$^1$, I.D.Karachentsev$^2$, S.N.Mitronova$^2$  and
W.K.Huchtmeier$^3$} \\

\bigskip

{\em $^1$Astronomical Observatory, Kiev National University, Observatorna ul.,
Kiev, 304053 Ukraine \\
$^2$Special Astrophysical Observatory, Russian Academy of Sciences,
Nizhnii Arkhyz, Karachai-Cherkessian Republic, 357147 Russia\\
$^3$Max Planck Institut f\"{u}r Radioastronomie, Auf dem H\"{u}gel 69,
D-53121 Bonn, Germany}
\end{center}

{\large \bf Abstract}\\
We analyze the peculiar velocity field for 2400 flat spiral galaxies
selected from an infrared sky survey (2MFGC). The distances to the
galaxies have been determined from the Tully--Fisher relation in
the photometric $J$ band with a dispersion of 0.$^m$45. The bulk motion
of this sample relative to the cosmic microwave background (3K) frame
has an amplitude of 199$\pm$37 km s$^{-1}$ in the direction
l = 290$^{\circ}\pm 11^{\circ} , b = +1^{\circ}\pm 9^{\circ}$.
The amplitude of the dipole motion tends to decrease with distance
in accordance with the expected convergence of bulk flows in the
3K frame. We believe that external massive attractors similar to
the Shapley cluster concentration are responsible for  $\sim$60\% of the
local flow velocity in the $z$ = 0.03 volume.

PACS numbers : N.98

DOI: 10.1134/S1063773706010014

{\bf Key words : galaxies, Tully--Fisher relation, large-scale motions.}

\bigskip

\section{INTRODUCTION}
 Studying the bulk flows as the deviations from uniform Hubble expansion
produced by large-scale gravitational effects is of particular interest
in testing various cosmological models (Peebles 1980). The peculiar
velocity of a galaxy is a measure of these deviations; apart from the
Hubble distance determination, an independent distance determination
is required to determine this velocity. The Tully--Fisher (TF)
relation (Tully and Fisher 1977) and the Faber--Jackson (FJ)
relation (Faber and Jackson 1976) together with its improved
version, the Fundamental Plane (FP) method, remain the main
popular methods for determining the distances to spiral and elliptical
galaxies, respectively. The accuracy of determining individual distances
by these methods is  $\sim$20\%, and the distances at which they
are applicable
can reach several hundred Mpc. In recent years, more accurate distance
determination methods where the absolute magnitudes of type-Ia
supernovae (SN Ia) (Riess et al. 1997) or the surface-brightness
fluctuations (SBF) of galaxies (Tonry et al. 2000) are the distance
indicators have been developed. Distances of several hundred Mpc
can be estimated from SN Ia to within 8\% (unfortunately, the number
of measured
supernovae is small). Distances as large as several tens of Mpc can
be estimated by the SBF method with an accuracy of 10--20\%.

A huge
body of observational data (see, e.g., Willick (2000) Courteau and
Dekel (2001), Zaroubi (2002) and references therein) pertaining
separately to E and S galaxies, field galaxies, cluster members,
etc. has been accumulated over 20 years of active studies of non-Hubble
flows. The works of different groups also differ by the depth of the
samples studied and the peculiarities of the sky distribution of
galaxies. The magnitude of the velocity and the direction of the
bulk motion can be determined by measuring the radial peculiar
velocities of galaxies. For the standard $\Lambda$CDM model, the bulk velocity
in the cosmic microwave background (3K) frame is expected to approach
zero as the volume under consideration increases. Therefore, measuring
the dipole component on various scales is needed to determine the
volume in which the flow converges. At present, there is no contradiction
between the velocities ($\sim$ 220 km s$^{-1}$ ) and the directions of the bulk
motions
of galaxies estimated by different authors on a scale of  $\sim$60 Mpc. At this
depth, the data for E and S galaxies, field galaxies, and clusters are in
good agreement, within the error limits, irrespective of the
distance determination method (see Table 1 in the review by Zaroubi
(2002)).

Two groups of results have been obtained at distances of
100--150 Mpc. Using 522 spiral galaxies in Abell clusters, Dale et al.
 (1999) determined the magnitude of the bulk velocity,
V = 75$\pm$92 km s$^{-1}$, and the apex position, $l = 282^{\circ}$ and
$b = 25^{\circ} (r\simeq 150$ Mpc), by the TF method.
Using 85 SN Ia, Riess et al. (1997) derived the apex position, $l=289^{\circ}$
and $b = -2^{\circ}$ , at a magnitude of the velocity that differs only
slightly from zero. The results obtained by Colless et al. (2001) using
the FP method apply to distant E galaxies in clusters located at distances
of $H_0 r = 6000 - 15000$ km s$^{-1}$  (EFAR) toward Hercules--Corona
Borealis and Perseus--Pisces--Cetus. The authors found the bulk velocity
in the volumes studied to be consistent with the hypothesis about its zero
value at a 5\% significance level. The above results are in good agreement:
on scales of 100--150 Mpc, the bulk velocity is 0--200 km s$^{-1}$ and
the apex positions coincide, within the error limits, and are close to
the direction of the excess of point sources in the sky from the IRAS
PSCz catalog (Saunders et al. 2000). It is important that the convergence
of bulk flows in this group of results was obtained for different types
of objects and using different distance determination methods.

The second
group of results is characterized by a high magnitude of the bulk velocity
and a significant spread in apex positions. Based on photometry for the
brightest galaxies in clusters, Lauer and Postman (1994) determined the
parameters of the motion relative to the 3K frame for 119 Abell clusters
of galaxies with velocities up to 15 000 km s$^{-1}$ (LP): $V = 689\pm178$
km s$^{-1}$ , $l = 343^{\circ}$, $b = 52^{\circ}\pm23^{\circ}$ .
Willick (1999) found the parameters of the bulk motion for 15
Abell clusters of galaxies located at a distance of 120 Mpc,
$V=720\pm280$ km s$^{-1}$, $l=272^{\circ}, b=10^{\circ}$
(using the TF method). Hudson et al. (2004) found the parameters
of the bulk motion in the 3K frame for 56 Abell clusters (the SMAC
sample) up to $H_0 r=6300$ km s$^{-1}$ with an effective sample
depth of $H_0r_e = 6300$ km s-1 : $V=687\pm 203$ km s$^{-1}$,
$l=260^{\circ}\pm13^{\circ}, b=0^{\circ}\pm11^{\circ}$ . The authors
believe that, apart from the Shapley Concentration, even more distant
 attractors must provide such a high velocity. Hudson (2003) attempted
to reconcile the conflicting data at depths from 6600 to 11000 km s$^{-1}$
by assuming that the sample sparseness and small size lead to large errors
in the peculiar velocities. If the LP sample is excluded, then the results
with high and low bulk velocities can be reconciled at a 2$\sigma$ level. The
sample that combines the SMAC survey with other surveys yields
a bulk flow with a velocity of 350$\pm$80 km s$^{-1}$ toward
$l = 288^{\circ}$ , $b = 8^{\circ}$ (Hudson 2003).

What the converge
scale is remains an open question until new observational data based
on much more complete and homogeneous samples are obtained. There are
advantages and disadvantages in using clusters of galaxies to study the
bulk motions. The distance determination for a cluster is based on
averaging the distances to its individual members, which reduces the
error. However, the number of measured galaxies in clusters is generally
small. In addition, including the cluster field galaxies remains a
possibility. The number of clusters measured by various authors ranges
from 20 to 100. Even if the cluster samples are distributed more or
less uniformly over the sky, they are very sparse due to the small size.
The measurement errors of the peculiar velocities for field galaxies are
fairly large and increase with distance and sample incompleteness. This
shortcoming is compensated for by the possibility of increasing the
sample to hundreds or even thousands of galaxies. The Mark III catalog
(Dekel et al. 1999) is an example of combining several catalogs of peculiar
velocities to determine the parameters of the bulk motion. This catalog
includes  $\sim$3000 E and S galaxies located at $R <$ 60 Mpc. The
Mark III galaxies move relative to the 3K frame with a bulk velocity
of $V = 370$ km s$^{-1}$ toward $l = 305^{\circ}$, $b = 14^{\circ}$.
This velocity seems fairly high compared to $V$ obtained for other
samples located in the same volume. The authors believe that an
uncertainty in the mutual calibration of the heterogeneous samples
may be responsible for the discrepancy.

Thus, an extensive deep homogeneous sample of galaxies uniformly and
densely distributed over the sky is needed for a more definite solution
of the question regarding the convergence scale of bulk flows. The
edge-on flat galaxy catalog, RFGC (Karachentsev et al. 1999), satisfies
these conditions. The RFGC includes 4236 galaxies with optical
angular diameters $a > 0.6^{\prime}$ and apparent axial ratios
$a/b > 7$. The RFGC is currently the most uniform all-sky catalog
of edge-on late-type spiral galaxies. About 3000 objects in the
catalog have fairly reliable $J, H, K$ photometry in the 2MASS survey
(Cutri and Skrutski 1998), and  $\sim$1200 objects have estimates of the
radial velocities $V_h$ and the hydrogen line widths $W_{50}$. We constructed
the 2MASS--TF relations for a sample of 1141 galaxies and, after
cleaning, determined the parameters of the bulk motion for galaxies
with $V_{3K} < 12000$ km s$^{-1}$ using 921 galaxies:
$V = 226\pm62$ km s$^{-1}$, $l = 295^{\circ}\pm16^{\circ}$,
and $b = -2^{\circ}\pm13^{\circ}$ (Kudrya et al. 2003).

Supplementing the 2MASS--RFGC sample to study the bulk flows suggests
performing new measurements of $V_h$ and $W_{50}$ for RFGC galaxies. In
particular, this program has been implemented with the 100-m Effelsberg
radio telescope by Huchtmeier et al. (2005) and Mitronova et al. (2005)
since 2001. Compiling a deeper homogeneous sample of late spirals based
on the 2MASS catalog (2MFGC) is a fundamentally different possibility
(Mitronova et al. 2004).

\section{THE 2MFGC CATALOG}
 The successful application of the infrared (IR) Tully--Fisher relation
to RFGC flat galaxies to determine the parameters of the bulk motion
(Karachentsev et al. 2002; Kudrya et al. 2003) was a justification for
compiling the new 2MFGC catalog. The 2MASS survey is known to have a
low sensitivity with respect to late-type galaxies (Jarrett 2000),
since the periphery of the disks of spiral galaxies is unseen on isophotes
fainter than $K_s = 20^m$.

 Studying the near-IR properties of RFGC galaxies
and comparing them with optical data allowed the selection criteria to be
worked out when compiling the new flat galaxy catalog based on 2MASS.
Thus, for example, comparison of the IR and optical diameters of RFGC
galaxies showed that the 2MASS diameters are, on average, half
the standard optical radii. For RFGC galaxies, the optical axial ratios
$a/b$ cover the range (7--21) with a median of 8.6, while the corresponding
IR axial ratios lie within the range (1--10) with a median of 4.1
(Karachentsev et al. 2002).

The principles of selecting galaxies for the new catalog from the original
2MASS--XSC catalog were described in detail by Mitronova et al. (2004).
The main parameter in the selection was the apparent axial ratio given
in the XSC (Extended Sources Catalog). The all-sky 2MFGC disk galaxy
catalog (Mitronova et al. 2004) contains IR photometry and identifications
in the LEDA and NED databases for 18020 objects with axial ratios
$a/b\geq  3$ in the XSC--2MASS. Analysis of the distribution of galaxies
in axial ratio in the new catalog yields a median value of $a/b = 4$,
which is close to the median value of $a/b = 4.1$ for the RFGC galaxies
identified in the XSC--2MASS. The two catalogs are similar in morphological
composition and contain  $\sim$80\% of the late-type spiral galaxies.

Below, we
use the following photometric parameters from the 2MFGC: $J_{fe}$ ,
the elliptical Kron magnitude; $r_{fe}$ , the elliptical Kron radius
(in arcseconds) in the K band; $a/b$, the axial ratio averaged over all $J$ ,
 $H$ , and $K$ bands; and IC, the concentration index (the ratio of the
radii within which 3/4 and 1/4 of the light from the galaxy is concentrated).

\section{A SAMPLE FOR DETERMINING THE PARAMETERS OF THE BULK MOTION}

 Compiling a sample of 2MFGC galaxies with estimates of the heliocentric
radial velocity $V_h$ , the HI line width $W_{50}$ , $W_{20}$ , or the
maximum rotational velocity $V_m$ are the subject of a separate paper.
 The radial velocities are known for 5536 of
the 18 020 2MFGC galaxies, and estimates of $W_{50}$ , $W_{20}$ ,
or $V_m$ are also available for 2765 objects. The widths $W_{50}$
 (N=2276) and $W_{20}$ (N=1825) are represented most completely in the
sample. Both $W_{50}$ and $W_{20}$ estimates are available for 1740
sample galaxies. Optical measurements of $V_m$ are available for 445 galaxies;
32 of them also have $W_{50}$ estimates. To choose the basic width,
$W_{50}$ or $W_{20}$ , in determining the parameters of the bulk motion,
we first constructed the standard twoparameter TF relations for both widths.
We take the TF relation in the form
$$M=C_1+C_2 \log W^c.\eqno(1)$$
We calculate the absolute magnitude from the apparent magnitude
$J_{fe}$ in the standard way,
			     $$M=J_{fe}-25-5\log r,\eqno(2)$$
by first estimating the photometric distance $r$ [Mpc] using the post-Hubble
relation
$$r=V_{3K}/H_0 \{1-(q_0-1)V_{3K}/2c \},\eqno(3)$$
where the radial velocity $V_{3K}$ in the CMB frame is calculated from the
heliocentric radial velocities $V_h = cz$ using the parameters of the solar
motion relative to the background from Kogut et al. (1993). We take the
Hubble constant $H_0 = 75$ km s$^{-1}$ Mpc$^{-1}$ and the deceleration
parameter $q_0 = -0.55$ in accordance with the cold dark matter (CDM)
cosmological model and the cosmological constant
$(\Omega_m=0.3,\quad \Omega_{\Lambda}=0.7)$.

We correct the widths for the cosmological expansion,
$$W^c=W/(1+z),\eqno(4)$$
 by abandoning the correction for turbulence, which was applied in our
previous paper (Kudrya et al. 2003), since it turned out to have no effect
on the bulk velocity parameters. We determine the coefficients of relation
(1) by least squares with equal weights.

We applied mainly the same data cleaning technique that was used
previously (Kudrya et al. 2003) by excluding galaxies with deviations
of more than $3\sigma_{TF}$ in the TF diagram and with individual peculiar
velocities in the 3K frame $V_{pec} > 3000$ km s$^{-1}$ . In contrast to
our previous paper (Kudrya et al. (2003), we did not exclude distance
galaxies ($V_{3K}>18000$ km s$^{-1}$, $H_0r>18000$ km s$^{-1}$) since there
were only a few such objects. After the cleaning, we obtained samples
of N = 1604 galaxies with $W_{20}$ and N = 1972 galaxies with $W_{50}$.
The
\begin{table*}[htbp]
\caption{Parameters of the Tully--Fisher relation for various samples}
\begin{tabular}{|l|l|l|l|l|r|}
 \hline N
&\multicolumn{1}{|c|}{Sample}
&\multicolumn{1}{|c|}{$\sigma_{TF}$}
& \multicolumn{1}{|c|}{$C_1$}
& \multicolumn{1}{|c|}{$C_2$}
& \multicolumn{1}{|c|}{$\sigma_{V}$}
 \\
 \hline 1&$W_{20},N1825$& 1.01& $-5.42\pm0.35$&$ -6.46\pm0.14$& 1925\\
2&$W_{50},N2276$& 0.97& $-7.14\pm0.28$&$ -5.86\pm0.11$& 1927\\
3&$W_{20},N1604$& 0.55& $-1.14\pm0.23$&$ -8.15\pm0.09$& 992\\
4&$W_{50},N1972$& 0.49& $-2.10\pm0.18$&$ -7.86\pm0.07$& 974\\
5&$W_{50}+V_m,N2684$& 0.92& $-6.80\pm0.25$&$ -6.01\pm0.10$& 1921\\
6&$W_{50}+V_m,N2333$& 0.49& $-2.55\pm0.17$&$ -7.69\pm0.07$& 1015\\
7&$W_{50}+V_m+W_{20},N2765$& 0.94& $-7.11\pm0.25$&$ -5.88\pm0.10$& 1930\\8&$W_{50}+V_m+W_{20},\quad N=2395$& 0.49& $-2.68\pm0.17$&$ -7.64\pm0.07$& 1016\\

 \hline
\end{tabular}
\end{table*}

\noindent results of our calculations are
presented in the first two rows of Table 1 for the complete samples of
N = 1825 galaxies with $W_{20}$ and N = 2276 galaxies with $W_{50}$ and
in the third and fourth rows for the cleaned samples.

In Table 1, $\sigma_{TF}$ denotes the standard deviation (in magnitudes)
in relation (1), $C_1$ and $C_2$ are the coefficients in (1), and $\sigma_V$
 is the standard deviation (in km s$^{-1}$ ) in peculiar velocity space.
In both cases, the scatter in the TF diagram is seen to be smaller for the
widths $W_{50}$. Therefore, we take $W_{50}$ as the basic width and
then successively add the data with $V_m$ and $W_{20}$ to the sample with
known $W_{50}$.

To pass from $V_m$ to $W_{50}$ , we used the relation
				$$W_{50}=2V_m+(27.4\pm7.8),\eqno(5)$$
derived from 32 galaxies with measured $W_{50}$ and $V_m$. In this way,
we enlarged the sample by 408 objects, thereby bringing it to N = 2684.

The calculated parameters of the TF relation for the complete N = 2684
sample and for the N = 2333 sample cleaned using the adopted technique
are given in the fifth and sixth rows of Table 1. We see that adding
the data with $V_m$ slightly reduced the scatter in the TF relation for
the uncleaned data and left the scatter for the cleaned data unchanged.

Our sample contains 1740 objects with measured $W_{50}$ and $W_{20}$.
The relation between these widths is shown in Fig. 1.

After excluding
the 68 galaxies that deviate from the linear relation by more than $3\sigma$,
we obtain the relation
$$W_{50}=W_{20}-(19.8\pm0.5).\eqno(6)$$

The hypothesis that the coefficient of the linear correlation between
the widths is equal to unity is confirmed by the Fisher test for an
orthogonal regression. Adding the data with

\begin{figure}[h]
\centerline{\includegraphics[angle=0, width=7cm]{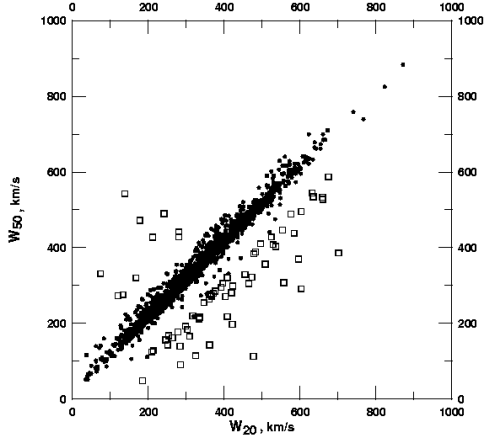}}
\caption{Relation between $W_{50}$ and $W_{20}$.
The squares mark 68 galaxies that deviate
from the linear relation by more than 3$\sigma$.}
\end{figure}

\begin{figure}[h]
\centerline{\includegraphics[angle=0, width=14cm]{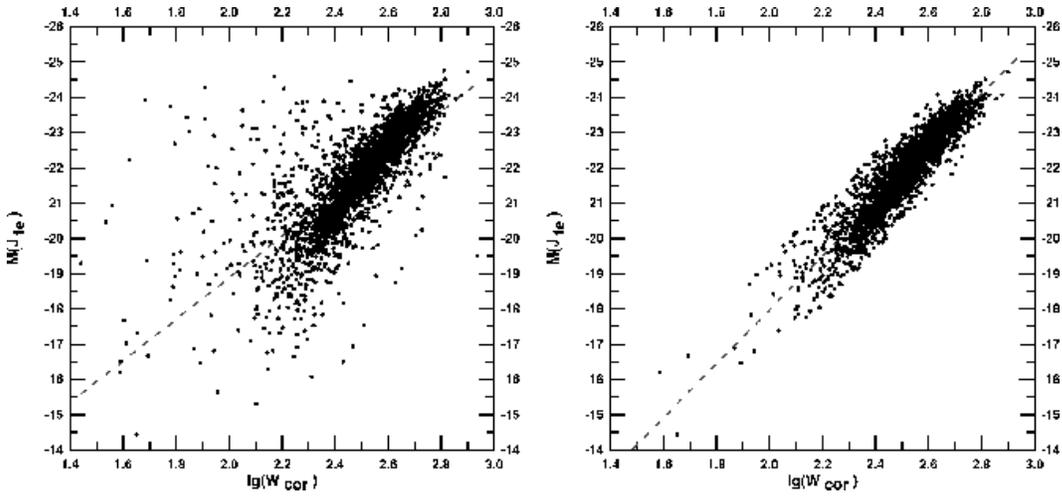}}
\caption{Tully--Fisher relations for the initial (a)
and cleaned (b) samples.}
\end{figure}

$W_{20}$ enlarged the sample
by 81 galaxies. Parameters of the TF relation (1) for the complete,
N = 2765, and cleaned, N = 2395, samples are given in the seventh and
eighth rows of Table~1.

The Tully--Fisher relations for the initial and cleaned samples are shown
in Fig. 2. The dashed lines in this figure correspond to the linear
regression constructed by least squares.

We give Fig. 2a to illustrate
the quality of the initial observational data. The region where most of
the galaxies are concentrated is clearly seen, but an appreciable fraction
of them (13\%) exhibit a large scatter in the diagram. As we see from Fig. 1,
 this scatter is partly (by about 1/3) attributable to the low quality of
the HI spectra. Comparison of Figs.~2a and 2b shows that our data
cleaning criteria are efficient.

In the next section, we determine the parameters of the bulk motion for
the cleaned N = 2395 sample.

Figure 3 presents the sky distribution of 2395 galaxies in Galactic
coordinates. The galaxies with measured $W_{50}$ (N = 1977),
$V_m$ (N = 353), and $W_{20}$ (N = 65) are marked by filled circles,
squares, and crosses, respectively. As we see, combining all of the data
improves appreciably the uniformity of the sky distribution.

\section{PARAMETERS OF THE BULK MOTION FOR 2MFGC GALAXIES}
\subsection{ Basic Formulas and General Results}
 We used a simple two-parameter TF relation to compile our main sample
of galaxies. To calculate
the parameters of the bulk motion, we generalize relation (1) by including
additional regressors. By an exhaustive search for various parameters of
galaxies, we found that three regressors are most significant according
to the Fisher test: the radius $r_{fe}$ , the axial ratio $a/b$, and the
concentration index IC. This linear relation is taken in the form
$$M=C_1+C_2 \cdot \log W^c + C_3 \cdot \log r_{fe}+C_4 \cdot a/b+C_5 \cdot IC.\eqno(7)$$

In this case, formulas (2)--(6) remain valid. We calculate the individual
peculiar velocity in the post
Hubble approximation,
$$V_{pec}=V_{3K}- H_0r\{1+(q_0-1)H_0r/2c\}.\eqno(8)$$

We use the set of peculiar velocities to calculate the orthogonal
components of the dipole bulk velocity $V$,
$$V_{pec,i}=\vec{V} \cdot \vec{e}_i+ \Delta V_i\eqno(9)$$
 by minimizing the sum of the squares of the ``noise'' peculiar velocity
component $V_i$ ($i$ is the galaxy number in the sample). Here,
$\vec{e}_i=(\cos l_i \cos b_i, \sin l_i \cos b_i, \sin b_i)$ is a unit
vector of direction of
\begin{figure}[h]
\centerline{\includegraphics[angle=-90, width=14cm]{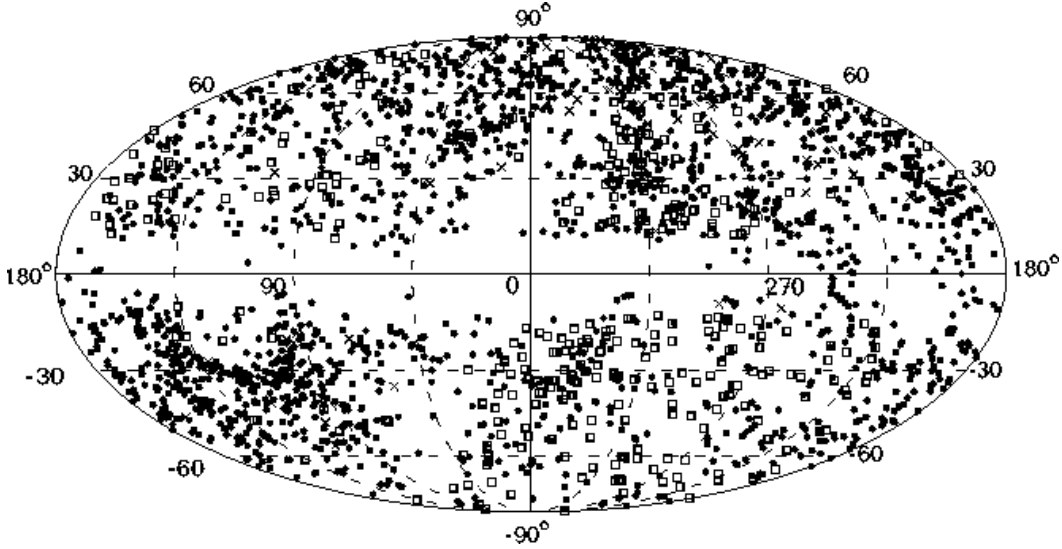}}
\caption{Distribution of 2395 2MFGC galaxies in Galactic
coordinates. Galaxies with measured $W_{50}$ , $V_m$ , and $W_{20}$
are denoted by filled circles, squares, and crosses, respectively.}
\end{figure}
galaxy $i$ in a reference frame associated with the Galactic coordinates
$l$ and $b$. We calculated the errors in the magnitude of the bulk
velocity $V$ and the apex position $l, b$ as follows. First, we determined
the diagonal elements $B_{VV}$ , $B_{ll}$ , and $B_{bb}$ of the covariance
matrix {\bf B} in the basis \{$\vec{e}_V, \vec{e}_l, \vec{e}_b$\}
and then
$\Delta V = (B_{VV})^{1/2}, \quad \Delta l =
arctg\{ (B_{ll})^{1/2}/
V\}, \quad \Delta b = arctg \{(B_{bb} )^{1 / 2} / V\}. $

The calculated bulk velocity parameters for the
N = 2395 sample and for the samples of 2MFGC galaxies with the
additional restrictions $V_{3K} < 8000$ km s$^{-1}$ and
$V_{3K} < 10000$ km s$^{-1}$ are given in the first three rows of Table 2.
Here, $V_x , V_y , V_z$ denote the orthogonal components of the dipole
bulk velocity, $V$ is its magnitude (all in km s$^{-1}$ ), $l$ and $b$ are
the Galactic coordinates of the apex, and $F$ is the significance
coefficient according to the Fisher test for the bulk velocity vector.
Comparison of the first row in Table 2 and the eighth row in Table 1
shows that including the additional regressors reduces the scatter in
the TF diagram by 6\%.

To test the robustness of the results obtained and to illustrate
the significance of the post-Hubble corrections in formulas (3)
and (8), we calculated the parameters of the five-parameter regression
and the bulk velocity by taking a linear Hubble law. The results for
the N = 2395 sample are given in the fourth row of the table. We see
that the apex position changed only slightly, while the magnitude of
the bulk velocity increased by $\sim$ 5\%.
Our calculations for other samples also confirm that allowance for the
relativistic redshif-distance relation slightly reduces the estimated
magnitude of the bulk velocity.

We also checked how the sample cleaning
based on the five-parameter regression (7) affected the parameters of
the bulk motion. The galaxy selection criteria were the same as those
for the two-parameter regression. The results of our calculations for
the new N = 2377 sample are presented in the fifth row of Table 2.
Comparison of the data in rows 5 and 1 shows a tendency for the magnitude
 of the velocity to decrease for a more rigorous sample
	   cleaning. At the same

\begin{table*}[htbp]
\footnotesize
\caption{Parameters of the dipole bulk velocity}
\begin{tabular}{|l|l|l|r|r|c|r|r|l|l|c|}
 \hline $N$&
 \multicolumn{1}{|c|}{Sample}&
 \multicolumn{1}{|c|}{$\sigma_{TF}$}&
\multicolumn{1}{|c|}{$\sigma_V$}&
\multicolumn{1}{|c|}{$V_x$}&
\multicolumn{1}{|c|}{$V_y$}&
 \multicolumn{1}{|c|}{$V_z$}&
 \multicolumn{1}{|c|}{$V$}&
\multicolumn{1}{|c|}{$l$}&
 \multicolumn{1}{|c|}{$b$}&
\multicolumn{1}{|c|}{$F$}\\
 \hline 1 &All,N2395&0.46 &962& $109\pm$37
& $-196\pm$37 & $-20\pm$31 & $225\pm$36& $299\pm10$&$-5\pm8$&13\\
 \hline 2 &$V_{3K}<10000$, N2205&0.48&$ 914$& $125\pm$37
& $-201\pm$37 & $-25\pm$30 & $238\pm$36& $302\pm9$&$-6\pm7$&15\\
 \hline 3 &$V_{3K}<8000$,N1943&0.49& 856& $87\pm$37
& $-206\pm$37 & $-22\pm$30 & $225\pm$36& $295\pm10$&$-6\pm8$&14\\
 \hline 4 &LHL,N2395&0.46& 970& $110\pm$37
& $-206\pm$38 & $-28\pm$31  & $235\pm$37& $298\pm9$&$-7\pm7$&14\\
\hline 5 & N2377&0.45& 968& $67\pm$37
& $-187\pm$38 & $2\pm$31 & $199\pm$37& $290\pm11$&$1\pm9$&10\\

 \hline
\end{tabular}
\end{table*}

\noindent  time, the data in Table 2 indicate that the sought-for
values of $V , l$, and $b$ agree, within the error limits.
Figure 4 presents the depth dependence of the magnitude of the
bulk velocity (dots with bars, left vertical scale) and the
sample size to a given depth (open diamonds, right vertical scale).

We see that the magnitude of the velocity decreases for distances
from 3000 to 8000 km s$^{-1}$, as expected in the standard model
for the formation of structures in the Universe with the cosmological
constant and cold dark matter. The sample size increases roughly linearly
to $V_{3K}$ = 8000--10000 km s$^{-1}$ and then is rapidly saturated.
After 8000 km s$^{-1}$, the magnitude of the velocity (230--240
km s$^{-1}$) and the apex position (which is not given here)
do not change,
\begin{figure}
\centerline{\includegraphics[angle=0, width=7cm]{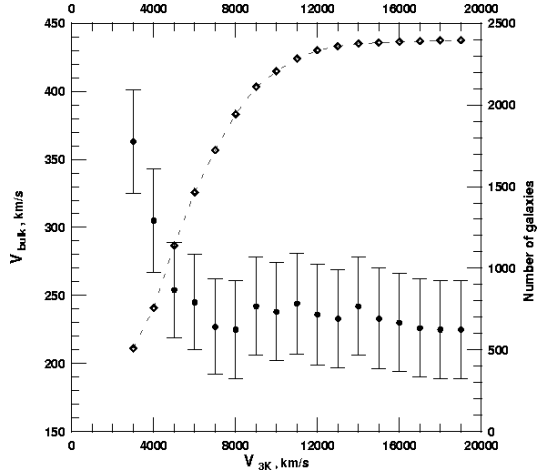}}
\caption{Magnitude of the bulk velocity (dots with bars,
left vertical scale) versus depth and the sample size to a
given depth (open diamonds, right vertical scale).}
\end{figure}
within the error limits. Thus, our results yield
the magnitude and direction of the bulk velocity for 2MFGC galaxies
to an effective depth of $\sim$ 10 000 km s$^{-1}$.

\subsection{Parameters of the Bulk Motion for Spherical Shells}
We also calculated the parameters of the bulk motion separately for four
subsamples of the complete N = 2395 sample by breaking it down into shells:
$0 < V_{3K} < 3000$ km s$^{-1}$, $S_1 ; 3000 < V_{3K} < 6000$ km s$^{-1}$ ,
$S_2; 6000 < V_{3K} < 9000$ km s$^{-1}, S_3;$
 and $9000 < V_{3K} < 20000$ km s$^{-1}, S_4$ . The volumes of the first,
second, third,
and fourth subsamples include, respectively, the Local Supercluster
$(l = 284^{\circ}$, $b = 74^{\circ}$), the Great Attractor
(Hydra, $l = 270^{\circ}, b = 26^{\circ}$; Cen 30+45, $l = 302^{\circ}$ ,
$b = 22^{\circ}$), the Coma supercluster ($l = 58^{\circ}, b = 88^{\circ}$),
and the Shapley Concentration ($l = 312^{\circ}, b = 31^{\circ}$).

The sky distributions of galaxies in the shells in Galactic
coordinates are shown in Fig. 5. The boundary of the shell
(in km s$^{-1}$ ) and the
number of galaxies in it are given in the upper left corner of each panel.
The regions of enhanced density corresponding to well-known superclusters
are clearly distinguished
in the individual panels. Table 3 presents the calculated dipole bulk
velocity. The parameters of the Tully--Fisher relation were determined
for each shell separately. The content of the columns is the same as
that in Table 3. As follows from these mutually independent data, the
apex position and the bulk velocity amplitude in each shell remain
approximately the same within (1--2)$\sigma$. The amplitude is at a
minimum, (157$\pm$45) km s$^{-1}$, in shell $S_2$ , where the Great
Attractor is located. We see no evidence of the flow toward the Virgo
cluster within the subsample $S_1$ with $V_{3K} < 3000$ km s$^{-1}$;
i.e., the entire volume of the Local Supercluster moves approximately
in the same direction as the remaining objects of our sample.

\subsection{Parameters of the Bulk Motion in the Frame of the Local Group}
As we noted above,  $\sim$13\% of the galaxies in the initial sample deviate
significantly from the linear TF regression (see Fig. 2a). The low
quality of the 2MASS photometry or the HI

\begin{figure}
\centerline{\includegraphics[angle=0, width=10cm]{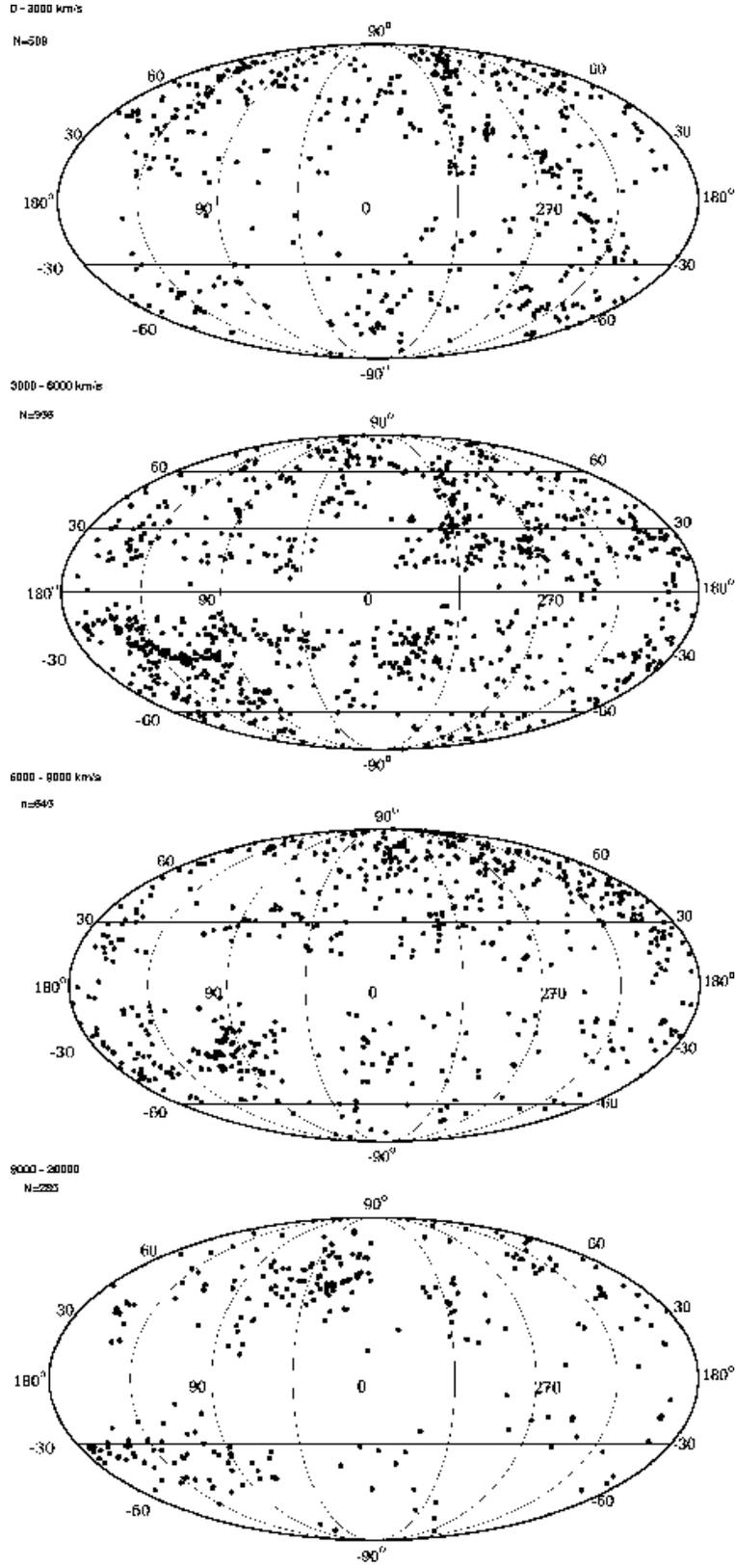}}
\caption{Sky distribution of galaxies in Galactic coordinates
in the shells. The boundaries of the shell (in km s$^{-1}$ )
and the number of galaxies in it are indicated in the upper
left corner of each panel.}
\end{figure}

\noindent spectra for these galaxies
is usually responsible for the deviations. Occasionally, the deviation
from the TF relation is caused by the peculiar structure of a galaxy or
the tidal perturbation from its close neighbors. We hoped to eliminate
these cases by applying the three sigma rule and restricting the
admissible absolute value of the peculiar velocity by the obvious
physical limit $V_{pec} < 3000$ km s$^{-1}$. As we see from relations
(1)--(3) and (8), the adopted conditions for eliminating ``bad'' galaxies
depend on the reference frame in which the distance and absolute magnitude
 of the galaxy are determined from its radial velocity. Therefore,
in principle, the dipole parameters can contain a systematic error caused
by the {\em a priori} choice of the 3K frame for estimating the
distances to galaxies.

According to Kogut et al. (1993), the Sun moves relative to the 3K
frame with a velocity of $V_{3K} = 369.5\pm3.0$ km s$^{-1}$ in the
direction $l = 264.4^{\circ}\pm0.3^{\circ}$,
$b = 48.4^{\circ}\pm0.5^{\circ}$, while according

\begin{table}[htbp]
\footnotesize
\caption{Parameters of the dipole bulk velocity for galaxies in the spherical shells}
\begin{tabular}{|l|l|l|r|r|l|c|l|l|l|c|}
 \hline $N$&
 \multicolumn{1}{|c|}{Sample}&
 \multicolumn{1}{|c|}{$\sigma_{TF}$}&
\multicolumn{1}{|c|}{$\sigma_V$}&
\multicolumn{1}{|c|}{$V_x$}&
\multicolumn{1}{|c|}{$V_y$}&
 \multicolumn{1}{|c|}{$V_z$}&
 \multicolumn{1}{|c|}{$V$}&
\multicolumn{1}{|c|}{$l$}&
 \multicolumn{1}{|c|}{$b$}&
\multicolumn{1}{|c|}{$F$}\\
 \hline 1 &$S_1$,N509&0.57& 438& $236\pm42$
& $-276\pm$36 & $27\pm$27 & $364\pm$38& $311\pm6$&$\phantom{m}4\pm4$&31\\
\hline 2 &$S_2$,N956&0.37& 738& $38\pm43$
& $-145\pm$44 & $-47\pm$41 & $157\pm$45& $285\pm16$&$-17\pm14$&4.4\\
\hline 3 &$S_3$,N645&0.30& 1003& $118\pm73$
& $-315\pm$83 & $-147\pm$61 & $367\pm$85& $291\pm11$&$-24\pm9$&6.4\\
\hline 4 &$S_4$,N285&0.25& 1187& $157\pm136$
& $-287\pm$139 & $-75\pm$109 & $335\pm$143& $299\pm22$&$-13\pm18$&1.8\\
 \hline
\end{tabular}
\end{table}

\noindent to Karachentsev and   Makarov (1996), the Local Group centroid has a velocity of $V_{3K}$ (LG) =
$634\pm12$ km s$^{-1}$ relative to the 3K frame in the direction
$l = 269^{\circ}\pm3^{\circ} , b = 28^{\circ}\pm1^{\circ}$ . The orthogonal
components of this velocity $V_x , V_y , V_z$ in the Galactic coordinate
system are ($-8, -559, 298$) km~s$^{-1}$ . To check the amplitude and
direction of the bulk motion for the possible presence of a systematic
error, we calculated the parameters of the dipole solution for 2MFGC
galaxies from their velocities relative to the Local Group centroid
and compared them with those obtained above in the 3K frame.

For N =
2404 galaxies that satisfied the three sigma rule and the condition
$V_{pec} < 3000$ km s$^{-1}$ in the LG frame, we obtained a standard
deviation relative to the two-parameter TF relation of
$\sigma_{TF} = 0.49^m$ or $\sigma_V = 986$ km s$^{-1}$ .
The bulk motion of these galaxies in the LG frame has a velocity of
$V = 372\pm33$ km s$^{-1}$ in the direction $l = 76^{\circ}\pm6^{\circ}$ ,
$b = -40^{\circ}\pm6^{\circ}$ ; the orthogonal components of this
velocity are . Since
$$\vec V_{3K}(2MFGC) =\vec V_{LG}(2MFGC) +\vec V_{3K}(LG),$$
 returning to the 3K frame, we obtain the orthogonal bulk velocity
components (+60, $-$284, +57) or the following new amplitude and
direction of the bulk motion: $V=296$ km s$^{-1}$, $l=282^{\circ}$,
$b=11^{\circ}$.
Comparison with the previous data (see row 1 in Table 2) shows
that the systematic difference in the parameters of the bulk motion
due to different galaxy elimination procedures lies within 2$\sigma$
of the random errors. However, it should be noted that the systematic
difference in $V_x , V_y , V_z$ is more significant in the close volumes
$S_1$ and $S_2$ .

\subsection{The Peculiar Velocity Field}
The distribution of the peculiar velocities for 2MFGC galaxies defined
by relations (7) and (8) is presented in Galactic coordinates in Fig. 6.
Galaxies with positive and negative peculiar velocities are indicated by
the open and filled circles, respectively. This
discrete distribution was averaged by a Gaussian filter with a 20$^{\circ}$
window. Equal mean peculiar velocities are indicated by the lines at
50 km s$^{-1}$ steps. The heavy line corresponds to a zero peculiar velocity;
negative $V_{pec}$ are indicated by the dashes.

As follows from these data,
the peculiar velocity distribution of 2MFGC galaxies is asymmetric: the
maximum positive isovelocity is +350 km s$^{-1}$, while the maximum
negative isovelocity is only 150 km s$^{-1}$. The $V_{pec}$ distribution
shows a distinct dipole pattern. The region of positive mean $V_{pec}$
appears doubly connected, with the primary peak being approximately in
the zone where the Hydra--Centaurus, Norma, A3627 clusters and the Shapley
Concentration of clusters are located. The secondary positive peak with
an amplitude of 150 km s$^{-1}$ is identified with the Abell cluster
A400 (170$^{\circ}$ , $-45^{\circ}$  ) and A539 (196$^{\circ}, -18^{\circ}$).
 The region of negative mean peculiar velocities forms three broad
shallow troughs with the coordinates of their centers (120$^{\circ}$,
+40$^{\circ}$), (80$^{\circ}, -30^{\circ}$), (200$^{\circ}$, +30$^{\circ}$).
The first two of them lie not far from the well-known Void in Bootes
and the Local Void. In general, the relief of the crests and troughs in
Fig. 6 agrees well with that obtained from a less representative sample
(N = 971) of RFGC galaxies with 2MASS photometry (Kudrya et al. 2003).
The reproducibility of general features in the $V_{pec}$ distribution
from two essentially independent observational samples gives hope that
the distribution of the gravitational potential in the local Universe
can be reconstructed from the available peculiar velocity map.

\section{DISCUSSION AND CONCLUSIONS}
 To estimate the representativeness and quality of the galaxy sample
used to determine the dipole parameters, we introduced the concept of
goodness, $G = (N/100)^{1/2}/\sigma_{TF}$ , where $N$ is the number of
galaxies in the sample, and TF is their dispersion (in magnitudes) in the TF
diagram (Kudrya et al. 2003). Of course, apart from $N$ and
$\sigma_{TF}$, the quality of
 the sample is also characterized by the uniformity with which it
covers the entire sky. However, allowance for the nonuniform distribution
of galaxies over the sky requires more complex approaches. At present,
there are only five samples whose goodness exceeds $G$ = 5. They are all
listed in Table 4 in order of increasing $G$. Columns 1, 2, and 3--5 give
the galaxy type, its goodness $G$, and the dipole amplitude and direction
in Galactic coordinates, respectively; the last column gives a reference
to the data source.

As follows from these data, the sample of 2MFGC galaxies being discussed
has the best goodness. The parameters of the 2MFGC dipole are in good
agreement with their weighted mean values for all five samples:
$V = 225\pm47$ km s$^{-1}$, $l = 295^{\circ}\pm3^{\circ} ,
b = +6^{\circ}\pm5^{\circ}$ . These amplitude and direction of the bulk
motion can be currently taken as the standard ones for the bulk flow of
galaxies relative to the CMB on a scale of  $\sim$100 Mpc. It should be noted
that this direction of the dipole is almost opposite to the direction
of the rotational velocity of the Sun relative to the Galactic center,
$V = 220$ km s$^{-1}$ , $l = 90^{\circ} , b = 0^{\circ}$ . For this
coincidental reason, the large-scale flow of galaxies had gone unnoticed
until the 1980s.

\begin{figure}
\centerline{\includegraphics[angle=0, width=14cm]{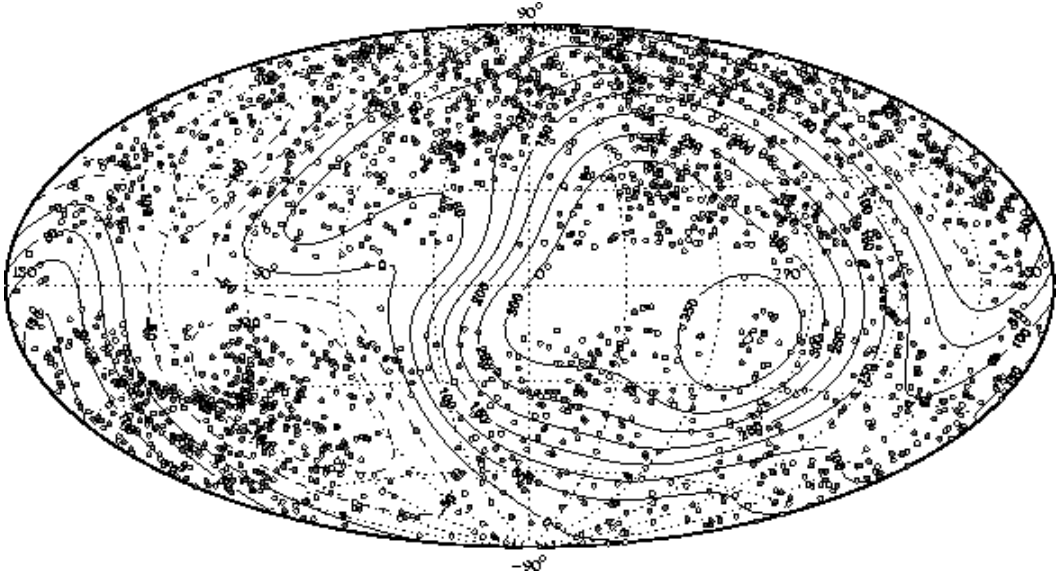}}
\caption{Peculiar velocity distribution for 2395 2MFGC galaxies in
Galactic coordinates. Galaxies with positive and negative peculiar
velocities are indicated by the open and filled circles, respectively.
The velocities of the galaxies were averaged by a Gaussian filter with
a 20 window. The lines of constant mean peculiar velocity are indicated
at 50 km s$^{-1}$ steps. The heavy line corresponds to a zero peculiar
velocity; negative $V_{pec}$ are indicated by the dashes.}
\end{figure}

Recall that we compiled and used catalogs of edge-on spiral galaxies to
analyze the bulk motions of galaxies. The selection of such galaxies by
their optical parameters (FGC) led to the following apex parameters:
$V = 300\pm75$ km s$^{-1} , l = 328^{\circ}\pm20^{\circ}$ ,
and $b = 78^{\circ}\pm158^{\circ}$ (Karachentsev et al. 1993). An
increase in the sample size in combination with the passage from optical
to IR photometry yielded new apex parameters: $V = 199\pm61$ km s$^{-1},
l = 301^{\circ}\pm18^{\circ}$, and $b = -2^{\circ}\pm15^{\circ}$
 (Kudrya et al. 2003). Using the new 2MFGC catalog compiled from
the 2MASS survey provided a more uniform distribution of galaxies over the
sky. The amplitude and direction of the bulk flow calculated here,
$V = 199\pm37$ km s$^{-1}$ , $l = 290^{\circ}\pm11^{\circ}$, and
$b = +1^{\circ}\pm9^{\circ}$, are in satisfactory agreement with our
previous dipole solutions (Karachentsev et al. 2000). Thus, our parameters
of the local large-scale flow turned out to be stable with respect to the
selection of galaxies by their optical or IR parameters, the use of the
optical or IR luminosities in the TF diagram. The dipole parameters
also change little if other regressors are included in the standard TF
relation between absolute magnitude and 21-cm line width: the axial
ratio, the surface brightness, the concentration index, and quadratic
combinations of these regressors. The manner of eliminating galaxies
by their peculiar velocity (in the 3K or LG frame) introduces a systematic
errors within 2$\sigma$ of the random errors.

\begin{table}[hbtp]
\caption{Goodness for various samples}
\begin{tabular}{|l|l|l|l|l|l|} \hline
Sample & G & $V$  & $l$& $b$& Reference \\ \hline
 Spirals in clusters & $6.0$ & 75 & 289 & 25 & Dale et al. (1999)\\
SNIa      &6.8    & 206   & 290  &  0 & Radburn-Smith et al. (2004) \\
  RFGC   & $ 7.4$ & 199   & 301  & --2 & Kudrya et al. (2003) \\
Mark III  & $8.6$ & 370   & 305  & 14 & Dekel et al. (1999) \\

2MFGC     &10.9   & 199  & 290 & 1  & This paper
\\ \hline
 Mean& ---     & 225  & 295 & 6  & ---

    \\ \hline
\end{tabular}
\end{table}

The apex of the bulk flow is located roughly in the same region as the
centroids of IRAS sources ($l = 258^{\circ} , b = +30^{\circ}$),
the centroid of 2MASS galaxies $(l = 278^{\circ}, b = +38^{\circ}$ ),
and the concentrations of rich Shapley clusters $(l = 315^{\circ} ,
b = +30^{\circ}$ ). If judged by the pattern of decrease in the flow
amplitude in Fig. 4 with linear scale of the Local Volume, then objects
(attractors) located farther than 50 Mpc could be responsible for $\sim$60\%
of the amplitude of the observed bulk motion of 2MFGC galaxies.

\bigskip

{\bf ACKNOWLEDGMENTS.} We wish to thank D. Makarov for help in preparing Fig. 6.
Here, we used data from the 2MASS survey, which is a joint project of the
Massachusetts University and the Infrared Data Processing and Analysis
Center sponsored by NASA and NSF. We used the LEDA database
(http://leda.univ-lyon1.fr). This study was supported by the DFG--RFBR grant
436RUS 113/701/0--1.

{}
\end{document}